\documentclass[preprint]{aastex}
\usepackage{amsbsy} 

\newcommand{\Msun}{M_\odot}

\slugcomment{Accepted to MNRAS}

\shorttitle{Cluster Evolution}
\shortauthors{Converse \& Stahler}

\begin{document}

\title{Star Clusters Under Stress:  Why Small Systems Cannot Dynamically Relax}

\author{Joseph M. Converse and Steven W. Stahler}
\affil{Astronomy Department, University of California,
    Berkeley, CA 94720}

\email{jconverse@astro.berkeley.edu}

\begin{abstract}

Utilizing a series of $N$-body simulations, we argue that gravitationally bound stellar clusters of modest population evolve very differently from the picture presented by classical dynamical relaxation theory. The system's most massive stars rapidly sink towards the center and form binary systems. These binaries efficiently heat the cluster, reversing any incipient core contraction and driving a subsequent phase of global expansion. Most previous theoretical studies demonstrating deep and persistent dynamical relaxation have either conflated the process with mass segregation, ignored three-body interactions, or else adopted the artificial assumption that all cluster members are single stars of identical mass.  In such a uniform-mass cluster, binary formation is greatly delayed, as we confirm here both numerically and analytically. The relative duration of core contraction and global expansion is effected by stellar evolution, which causes the most massive stars to die out before they form binaries.  In clusters of higher $N$, the epoch of dynamical relaxation lasts for progressively longer periods. By extrapolating our results to much larger populations, we can understand, at least qualitatively, why some globular clusters reach the point of true core collapse.

\end{abstract}

\keywords{open clusters and associations: general --- binaries: general --- stars: kinematics and dynamics, mass function}

\section{Introduction}

The dynamical evolution of gravitationally bound stellar clusters has been extensively studied for decades, and the basic theory is thought to be secure. Populous systems evolve, over many crossing times, through the processes known collectively as dynamical relaxation \citep[][Chapter~7]{bt08}. The inner core of the cluster contracts, effectively transferring energy to the outer halo, which expands as a result. Concurrently, stars of relatively high mass sink toward the cluster center. Theory predicts further that the interior contraction leads eventually to core collapse, a catastrophic rise in central density \citep{lw68}. As first suggested by \citet{hi75}, the runaway is halted when hard binaries form near the center and release energy through three-body encounters. Observations of globular clusters, which can be significantly older than their relaxation times, have confirmed these expectations beautifully. The surface brightness profiles of Milky Way globular clusters indicate that some 20~percent harbor collapsed cores \citep{dk86,cd89,t93}. Near the centers of many systems are X-ray binaries and blue stragglers \citep{b95}, both created at high stellar density, perhaps during the collapse phase.

Globular clusters have impressive populations (\hbox{$N\,\sim\,10^5\,-\,10^6$}), but are relatively rare and distant groups. Open clusters are sparser (\hbox{$N\,\sim\,10^2\,-\,10^3$}), but much more common and closer at hand, with over a thousand catalogued \citep{d02}; the sample is thought to be complete out to 2~kpc \citep{b01}. Ironically, their evolutionary status is much less clear. Half of open clusters disintegrate within \hbox{$2\times10^8~{\rm yr}$} after birth \citep{w74}, a span corresponding to at most a few initial relaxation times. Not surprisingly, there is little observational signature that relaxation has occurred, aside possibly from mass segregation, first found by \citet{vs60}. A small fraction of open clusters, located at the outskirts of the Galaxy, have survived for over 1~Gyr \citep{f95}. Even these ancient systems show no sign of core collapse. A prototypical example, M67, has a smooth surface density profile that, unlike post-collapse systems, is well fit by a King model \citep{bb05}; the system appears to be in the last stages of tidal disruption \citep{ds10}.

\citet{h05} performed $N$-body simulations to follow the evolution of M67 from infancy to its inferred age of 4~Gyr. Their preferred model began with 12,000 single stars and an equal number of binaries; only about 10~percent of these stars survived to the end. Even over the protracted time of the simulation, the cluster never exhibited classical dynamical relaxation. Instead, the central mass density rose slightly and then declined. \citet{h05} attributed this behavior to the binary-rich initial population. Hard binaries undergo superelastic encounters with other stars and effectively heat the cluster \citep{he75}, counteracting the outward energy transfer driving dynamical relaxation. Indeed, it has long been appreciated that the presence of even a few binaries can radically alter the evolution of the sparsest groups \citep{t87}. These findings prompt us to ask a more general question: Under what conditions does binary heating prevent significant core collapse?

In this paper, we begin to address this larger issue, utilizing our own suite of $N$-body simulations. The basic answer to our question is that {\it the presence of massive stars} is essential. These massive stars couple with others to form pairs that, through three-body interactions, frustrate core contraction relatively early, so that there is little or no rise of the central density. The system thereafter undergoes {\it global expansion}. Here, the stellar density falls everywhere.

However, there a mitigating factor in this scenerio - stellar evolution. In more populous clusters with longer relaxation times, the most massive stars die out. Binary heating is tamed, and does not effectively oppose core contraction until later in the cluster's evolution. As a result, this contraction proceeds to yield a higher density contrast. Such is the case in the globular clusters that have undergone true core collapse.

We elucidate these processes in a step-by-step fashion, beginning with very simple, highly idealized systems, and progressively adding more realistic features. In Section~2 below, we introduce an energy analysis that quantitatively distinguishes classical dynamical relaxation from global expansion. We apply this analysis to both single-mass systems and those with a more realistic stellar mass distribution; only in the latter does binary heating come into play. Section~3 describes in more detail the discontinuous manner in which binary heating actually operates, while Section~4 shows how stellar evolution lessens the effect. Finally, in Section~5, we discuss the implications of our results for both open clusters, which we have simulated more or less accurately, and globular clusters, which we cannot model directly.

\section{Cluster Energetics}

\subsection{Prelude: Single-Mass Models}

We have run a suite of N-body simulations, all employing the publicly available code Starlab \citep[Appendix B]{pz01}. Significantly for our purposes, the code uses no softening in the gravitational potential, so that the formation and dynamical interactions of binaries are followed accurately. As we did not have access to any special-purpose hardware, our simulations were limited by time constraints to \hbox{$N\,\lesssim\,10^4$}.

Let us first adopt the simplified, and assuredly unrealistic, assumption that all stars have identical mass. Many of the classic theoretical papers in stellar dynamics, as well as textbook accounts, have utilized such single-mass models. We assume the cluster starts out in virial equilibrium, with a mass density profile corresponding to an \hbox{$n\,=\,3$} polytrope. We recently found that this particular configuration best describes the very early state of the Pleiades, just after gas removal \citep[][hereafter Paper~I]{cs10}. We stress, however, that the precise initial state is of little consequence; the system loses memory of this state well within one relaxation time, as in the case of the Pleiades.

In this and our other simulations, we model only isolated systems, with no tidal gravitational field either from the Galaxy or from passing molecular clouds. More complete models should include such an external field, which eventually destroys all clusters. However, the presence of the field does not qualitatively alter our main conclusions regarding cluster evolution up to the point of dissolution.

Figure~1 shows the evolution of Lagrangian mass shells in a cluster with the representative population of \hbox{$N\,=\,4096$}. Here we display the temporal change of the shell radii, expressed as fractions of the cluster's initial virial radius, $r_v$. Each shell has the indicated value of $M_r/M_0$, where $M_r$ is the interior mass and $M_0$ the initial mass of the whole cluster. The time itself is normalized to the initial relaxation time $t_{\rm relax}$, for which we utilize equation~(1.37) in \citet{bt08}:
\begin{equation}
t_{\rm relax} \,\equiv\, {N\over{8\,{\rm ln}\,\Lambda}}\,\,t_{\rm cross} \,\,,
\end{equation}
where \hbox{$\ln \Lambda\,=\,\ln(0.4\,N$)}. Following standard practice \citep[e.g.,][Section~2.4]{pz98}, the crossing time $t_{\rm cross}$ is given by
\begin{equation}
t_{\rm cross} \,\equiv\, \left({{8\,r_v^3}\over{G\,M_0}}\right)^{1/2} \,\,.
\end{equation}
We need not choose values for $r_v$ or $t_{\rm cross}$ as long as we compare only nondimensional versions of all the relevant quantities \citep{hm86}.

As we see in the figure, interior shells contract, while those closer to the cluster boundary expand. This behavior is the hallmark of dynamical relaxation. In this plot, the steeply accelerating contraction that signifies core collapse is not present, simply because of the limited time range covered. \citet{m96}, who investigated single-mass models using a special-purpose (Grape-4) computer, found core collapse to occur after about 6 initial relaxation times. (See his Figure~1 plotting the central density for the 4k run, after noting that $t_{\rm relax}$ corresponds to 62 of his scaled N-body time units.) Our results, over a more restricted interval, are fully consistent with Makino's.

Other researchers, utilizing a variety of techniques, have verified through simulations that dynamical relaxation occurs in single-mass systems \citep[e.g.,][]{t95,b03}. Nevertheless, it is worth revisiting the basic energetics of the process. The cluster's total energy is conserved, so its dual contraction and expansion reflects energy {\it transfer} from the inside out. According to Figure~1, the shell with \hbox{$M_r/M_0\,=\,0.70$} grows only slowly. Thus, this shell lies just outside the core-halo boundary.

We arrive at the same conclusion by calculating directly the mean rate of energy transfer. Let $K_r$ be the total kinetic energy within mass $M_r$, and $\dot K_r$ the time derivative of this quantity. After finding the best-fit straight line to $K_r (t)$ over the full time span of the simulation (\hbox{$3\,\,t_{\rm relax}$}), we then calculate $\dot K_r$ as the slope of this line. Figure~2 shows $\dot K_r$ as a function of the mass fraction $M_r/M_0$. What we actually display is the nondimensional quantity $\dot k_r$, where   
\begin{equation}
{\dot k_r} \,\equiv\,{{{\dot K_r}\,t_{\rm relax}}\over K_i} \,\,,
\end{equation}
and where $K_i$ is the cluster's total initial kinetic energy.

The curve in Figure~2 has a small, central dip, a numerical artifact of the large scatter in $K_r (t)$ over this region containing relatively few stars. Thereafter, $\dot K_r$ rises, attains a maximum, and then monotonically declines. Within the rising portion of the curve, the kinetic energy \hbox{$\Delta K_r$} in a shell of thickness $\Delta M_r$ {\it increases} with 
time. That is,
\begin{eqnarray}
\Delta K_r \,&=&\, {{\partial\,{\phantom t}}\over{\partial\,t}}  
\left({{\partial\,K_r}\over{\partial\,M_r}}\,\,\Delta M_r\right) 
\Delta t \nonumber \\
 &=&\, {{\partial\,{\dot K_r}}\over{\partial\,M_r}}\,\Delta M_r\,\,\Delta t\\
 &>&\, 0 \nonumber \,\,. 
\end{eqnarray} 
A certain, interior region of the cluster is thus gaining kinetic energy. Self-gravitating systems have negative heat capacity. Thus, the increasing kinetic energy (and therefore temperature) of the inner core signifies a {\it decreasing} total energy.\footnote{This argument is only suggestive, as the gravitational potential energy of any interior region actually depends on the distribution of mass surrounding it.} Analogous reasoning shows that the region corresponding to the descending portion of the \hbox{${\dot K_r} - M_r$} curve is {\it gaining} total energy, and therefore comprises the halo, which receives its energy from the core. It is natural, therefore, to locate the core-halo boundary at the peak of the curve, i.e., where \hbox{${\dot K_r}\,=\,0$}. According to Figure~2, this boundary is at \hbox{$M_r/M_0\,\approx\,0.6$}, in agreement with the analysis of Figure~1.

\subsection{Models with a Realistic Stellar Mass Distribution}

We next eliminate the most egregious simplification in the model, the assumption of a uniform stellar mass. As has long been appreciated \citep[e.g.,][]{iw84,fm95}, relaxing this assumption has a profound effect on cluster evolution. We turn again to our recent study of the Pleiades (Paper~I), and use, as our stellar mass distribution, the one characterizing the cluster in its infancy, soon after gas dispersal. This distribution was a lognormal, joining smoothly onto a power law at higher masses. The full distribution is given in equation~(17) of Paper~I, with the parameter values listed in Table~I of the paper. Following that study, we take the minimum and maximum stellar masses to be $m_{\rm min} = 0.08$ and $m_{\rm max} = 10~\Msun$, respectively. (See Section 3.3 for reconsideration of the maximum mass.) Once again, we assume that the mass density profile of the cluster is that of an \hbox{$n\,=\,3$} polytrope. Cluster members are all single stars, whose masses are drawn randomly from the assumed stellar distribution. We set \hbox{$N\,=\,4096$}, and ignore both mass loss during stellar evolution and any tidal gravitational field.

The left panel of Figure~3 shows the evolution of Lagrangian mass shells, in a manner analogous to Figure~1. In this case, we note first that radii tend to exhibit more jitter in their evolution. This characteristic stems from the redistribution of stellar mass over the crossing time. Even after averaging over the jitter, interior mass shells do not monotonically contract, as they did before. These radii initially shrink. However, at some relatively early time \hbox{$t_b\,=\,0.37\,\,t_{\rm relax}$}, they reach a minimum and begin to expand. (We will later identify this time with binary formation; see below.) This expansion continues, with ups and downs, for the remainder of the simulation. Radii corresponding to \hbox{$M_r/M_0\,\gtrsim\,0.7$} expand from the start.

For the single-mass model, the shrinking of any interior radius unambiguously signifies that the average distance between stars is also diminishing in that region. In the present case, the interpretation of early contraction is complicated by the phenomenon of mass segregation. The mass of any star is drawn from the same distribution, regardless of that object's initial location in the cluster.\footnote{More precisely, there is no correlation between a star's mass and its energy; see Paper~I, Section~2.1.} Thus, there is no mass segregation initially. However, relatively massive stars quickly drift toward the center, under the influence of dynamical friction. As these stars accumulate, the radius of any region {\it of fixed mass} may shrink, even if the average interstellar spacing does not.

To illustrate this point graphically, the right panel of Figure~3 shows, for the same simulation, the evolution of radii containing a fixed {\it number fraction}, $N_r/N$, of the cluster. Here, $N_r$ is the interior number of stars. Since no mass segregation was imposed at the start, the Lagrangian mass shell with \hbox{$M_r/M_0\,=\,0.10$} initially has the same radius as the ``number shell'' with \hbox{$N_r/N\,=\,0.10$}. The radius of the former contracts at early times, but Figure~3b shows that the radius of the latter stays constant and later grows. The average interstellar separation within the volume is not shrinking.

Tracking the radii of Lagrangian mass shells is a widely employed technique for visualizing cluster evolution. Other authors who have studied relatively low-$N$ systems under similar assumptions have documented the early contraction of interior shells. This development is said to demonstrate core collapse \citep[e.g.,][]{gh97,h04}. The putative collapse occurs within one initial relaxation time, much earlier than in single-mass models, and ends before the central mass density has risen dramatically. However, contraction of inner mass shells may be due to mass segregation and thus it alone does not definitively show the occurrence core collapse.

In any cluster containing a range of stellar masses, any temporal increase in mass segregation obscures the physically distinct phenomenon of core contraction. The latter may still be occurring, and is in this case to a limited degree. Consider again Figure~3b, where we have added a deeply embedded, number shell corresponding to \hbox{$N_r/N\,=\,0.03$}. This shell {\it does} contract initially, although only by a relatively small amount before the turnaround at $t_b$. True dynamical relaxation occurs at the start, but the accompanying interior contraction is weak, and is soon aborted.

As in the single-mass model, calculation of the mean energy transfer rate adds physical insight. Figure~4 is an energy transfer profile before turnaround, i.e., over the interval \hbox{$0\,<\,t\,<t_b$}. The curve is similar to that in Figure~2. The rate ${\dot K_r}$ eventually rises smoothly, peaks a bit beyond \hbox{$M_r/M_0\,=\,0.5$}, and thereafter declines. Such a profile is again indicative of dynamical relaxation. Identifying the core-halo boundary with the peak of the \hbox{${\dot K_r}\,-\,M_r$} curve is consistent with the pattern of mass shell curves in Figure~3b. Thus, the radius corresponding to \hbox{$M_r/M_0\,=\,0.5$} initially contracts slightly, while that with \hbox{$M_r/M_0\,=\,0.7$} expands. 

Figure~5 displays the energy transfer profile after the turnaround. Here, the mean rate $\dot K_r$ is determined over the interval \hbox{$t_b\,<\,t\,<\,3\,\,t_{\rm relax}$}. The profile is now qualitatively different, and illustrates a distinct mode of cluster evolution. The kinetic energy in every mass shell falls with time. So does, therefore, the kinetic energy of the entire cluster. From the virial theorem, the cluster as a whole is gaining in {\it total} energy. This injection of energy accounts for the system's global expansion, as seen in all the radii of Figure~3 for \hbox{$t\,>\,t_b$}. The central engine driving the expansion is binary heating, as we verify shortly.

We have run analogous simulations for cluster populations ranging from \hbox{$N\,=\,512$} to 16,384. The upper limit was a practical one; the last case required three months on a desktop computer. All simulations gave qualitatively the same result. The cluster experiences an early, transient phase of dynamical relaxation. During this epoch, the central number density rises by only a modest amount, typically a factor of 2. The central mass density rises by about a factor of 10, with the larger increase reflecting the onset of mass segregation. In all cases the end of this early period coincides with the formation of the first long-lived binary system, with $t_b\,\approx\,0.3\,t_{\rm relax}$ for all $N$, consistent with the findings of \citet{pz02}. From this early epoch until the end of the simulation, the cluster undergoes global expansion. As we shall next see, even the brief, transient period of dynamical relaxation was itself an artifact that vanishes under more realistic initial conditions.

\subsection{The Example of the Pleiades}

One feature of our simplified cluster models is that they consist initially of only single stars. It is well known that most field stars of solar-type mass have at least one binary companion \citep{dm91}. The observational assessment of binarity in even the nearest open clusters is challenging, but the indication so far is that the fraction is comparable to the field-star 
value \citep[e.g.,][]{b97,ds10b}. Since, as we will see, binary heating plays a key role in dynamical evolution, we should try to gauge the influence of primordial pairs.

For this purpose, we may utilize our own simulated history of the Pleiades (Paper~I). Our initial state, another \hbox{$n\,=\,3$} polytrope, was that which evolved, over the 125~Myr age of the cluster, to a configuration most closely resembling the current one.\footnote{For most of the simulations in Paper~I, including those reviewed here, we ignored both stellar mass loss and the Galactic tidal field. Adding both effects had a negligible impact on the evolution up to the present age of the cluster (Paper~I, Section~3).} In a previous investigation \citep{cs08}, we found that 76~percent of the stellar systems today are binary. The best-fit initial state in Paper~I consisted essentially of {\it all} binaries, with the corresponding fraction being 95~percent. We endowed these binaries with a lognormal period distribution and a thermal distribution of eccentricities, reflecting both conditions in the field population and in the Pleiades itself \citep{dm91,b97}. In addition, the masses of the primary and secondary stars were correlated (see Section 2.1.2 of Paper~I).

Finally, the initial state had a finite degree of mass segregation, i.e., the masses and energies of stellar systems were also correlated. The reader is again referred to Paper~I for the detailed prescription. Mass segregation may be characterized quantitatively through the Gini coefficient \citep[][Section~4.2]{cs08}. This quantity measures how fast the cumulative mass increases outward relative to the cumulative number of systems. The initial state of the Pleiades had \hbox{$G\,\approx\,0.14$}. The initial number of stellar systems, both binary and single, was \hbox{$N\,=\,1215$}.

Figure~6 shows the evolution of mass- and number-shell radii. Note that the current age of the Pleiades corresponds to about 0.5 initial relaxation times. Thus, these plots span a significantly briefer interval than those in Figures 1 and 3. Bearing this fact in mind, we see that the curves are generally similar to those in Figure~3. After a brief initial plunge, the radii of mass shells with \hbox{$M_r/M_0\,\gtrsim\,0.3$} expand, while the \hbox{$M_r/M_0\,=\,0.1$} shell contracts, at least over this time. Number shells undergo an analogous, early contraction, and then either remain static (\hbox{$N_r/N\,=\,0.03$}) or expand. If binaries are energetically significant, why is the cluster evolution not radically altered? This is an important question, to which we shall return presently (see Section~3.2).

The early dips seen in all the curves of Figure~6 signify that the cluster as a whole initially contracts. This behavior is an artifact of our specific method for implementing mass segregation. As explained in Section~3 of Paper~I, the configuration starts out in precise virial equilibrium. However, the redistribution of higher stellar masses toward the center alters slightly the gravitational potential from that associated with an \hbox{$n\,=\,3$} polytrope. Over a period lasting about two crossing times (\hbox{$0.08\,\,t_{\rm relax}$}), the cluster ``bounces,'' and then settles into a configuration that evolves smoothly thereafter.

The bounce does not occur if we impose no mass segregation initially. In that case, both the mass density of stars and the gravitational potential correspond exactly to an \hbox{$n\,=\,3$} polytrope. Figure~7 shows results from such a simulation. In this ``Pleiades-like'' cluster, the initial state is identical to that in Figure~6, but without mass segregation. Over the time span covered ($0.5\,\,t_{\rm relax}$), number shells either remain static or expand (Figure~7b). The early contraction of the innermost shells seen in Figure~3 never occurs, due to the heating by primordial binaries. In summary, there is no evidence of classical dynamical relaxation; the cluster evolves purely through expansion. The radii of interior mass shells do contract (Figure~7a) as a result of increasing mass segregation; the Gini coefficient grows from 0 to 0.15 over this time (see Figure~8 of Paper~I).

Returning to the more realistic Pleiades simulation, it is again instructive to visualize the internal transport of energy. From our description thus far, there should be no core-halo boundary, identified by the peaks of the energy transfer profiles in Figures~2 and 4. Figure~8, which plots $\dot K_r$ as a function of $M_r$, bears out this expectation. Here, we have computed $\dot K_r$ by a linear fit over the full time span of the simulation. We see that $\dot K_r$ monotonically falls from zero to increasingly negative values. (Compare Figure~5 and the accompanying discussion.) The cluster as a whole is cooling down, and is therefore gaining in total energy. The Pleiades evolved to its present state through global expansion, not dynamical relaxation.

\section{The Role of Binaries}

\subsection{First Appearance}

Let us reconsider the highly idealized clusters with which we began - one with stars of identical mass, and the other with a continuous range of stellar masses. In both cases, the initial systems contained neither binaries nor higher-order multiple systems. The evolving, single-mass cluster spawned no new binaries over the duration of our simulation. However, \citet{m96} found, in his more extensive investigation of the single-mass model, that binaries do eventually form in the contracting interior, and that their heating reverses core collapse at \hbox{$t\,\approx\,6\,\,t_{\rm relax}$}. The core subsequently undergoes the gravothermal oscillations predicted by \citet{bs84} and \citet{g87} using fluid models with an internal energy source.
  
In our cluster with a realistic stellar mass distribution, the interior contraction ends much sooner, within a single initial relaxation time. Is this prompt reversal also due to binary heating? The answer is yes. We have confirmed that the turnaround at \hbox{$t\,=\,t_b$} coincides with the appearance of the first hard binary. Here, we remind the reader that a ``hard'' binary is one whose gravitational binding energy exceeds the average, center-of-mass kinetic energy of all other stellar systems. It is only such pairs that donate energy to neighboring stars during a close flyby, and thereby become even harder. This is the essence of binary heating \citep{he75}.

Why do binaries form so much earlier in this cluster than in the single-mass model? Closer inspection reveals that these new systems are comprised of stars that are appreciably more massive than the average cluster member. This fact is readily understood in a qualitative sense. In clusters with no initial mass segregation, the relatively massive stars promptly sink to the center. Once in close proximity, these objects have a stronger mutual attraction than other cluster members, and are thus more prone to forming binaries.

In more detail, a gravitationally bound pair of such stars can only form by giving energy to a third star. Binary formation is thus a three-body process. In the traditional analysis of cluster evolution based on single-mass models, three-body encounters throughout the bulk of the system are considered too rare to be of significance. \citet[][p.~558]{bt08} show that $t_b^\ast$, the time for the first binary to form via this route, is much longer than $t_{\rm relax}$. Specifically, they estimate that
\begin{equation}
{t_b^\ast\over{t_{\rm relax}}} \,\sim\, 10\,N\,{\rm ln}\,N \,\,.
\end{equation}   
Our superscript on $t_b^\ast$ emphasizes that this time pertains to the highly specialized case of equal-mass stars. The derivation of equation~(5) assumes that the binary-forming stars reside in a region of average density. This assumption breaks down if the interactions occur in a deeply collapsing core. Three-body interactions {\it can} proceed here efficiently 
\citep[e.g.,][]{he84}. However, the process is too slow in regions where the density is not greatly enhanced.\footnote{Although binaries could, in principle, form via three-body interactions within globular clusters, those that eventually arrest core collapse are actually extremely tight systems created earlier by tidal capture \citep{f75}.} 

The situation changes dramatically once the cluster is endowed with a distribution of stellar masses. Here, binaries can form even where the density is close to the average. We may demonstrate this fact through a slight alteration of the heuristic derivation for $t_b^\ast$ given in \citet{bt08}. Suppose that stars require a minimum mass $m$ to be part of a binary. The time $\Delta t$ for a given one of these objects to come within distance $b$ of another with comparable mass is
\begin{equation}
\Delta t \,\sim\, \left(f_m\,n\,b^2\,\sigma\right)^{-1} \,\,.
\end{equation} 
Here $f_m$ is the number fraction of such stars, $n$ is the average cluster number density, and $\sigma$ the velocity dispersion. During this encounter, there is a probability \hbox{$p\,\sim\,4 \pi n\,b^3 / 3$} that a third star will also be within the interaction distance $b$. This star can have the average mass \hbox{$\langle m\rangle\,\equiv\,M_0/N$}. Thus, the time for the original star to suffer a binary-forming triple encounter is about \hbox{$\Delta t/p\,=\,3 / \left({4\,\pi\,f_m\,n^2\,b^5\,\sigma}\right)$}. There are $f_m\,N$ such stars in the cluster. The time for {\it any} such star to form a binary is
\begin{equation}
t_b \,\sim\,{3\over{4\,\pi\,N\,f_m^2\,n^2\,b^5\,\sigma}} \,\,.
\end{equation}

In order for a hard binary to form, the gravitational potential energy of the binary must be equal to or greater than the average kinetic energy in the cluster:
\begin{equation}
{{G\,m^2}\over b} \,\sim\, \langle m\rangle\,\sigma^2 \,\,. 
\end{equation} 
Thus,
\begin{equation}
t_b \,\sim\, {{3\,\sigma^9\,\langle m\rangle^5}\over
             {4\,\pi\,N\,f_m^2\,n^2\,G^5\,m^{10}}} \,\,.
\end{equation}
From the virial theorem, \hbox{$\sigma^2\,\sim\,G\,N\,\langle m\rangle/r_v$},
where $r_v$ is the cluster's virial radius. Using this expression along with the approximation that \hbox{$n \sim 3 N / \left(4 \pi r_v^3\right)$} we find
\begin{equation}
t_b \,\sim\, {{4\,\pi\,N^{3/2}\,r_v^{3/2}}\over
 {3\,f_m^2\,G^{1/2}\,\langle m\rangle^{1/2}}}
\left({\langle m\rangle\over m}\right)^{10} \,\,.
\end{equation} 
Now the relaxation time from equation~(1) may be approximated as
\begin{equation}
t_{\rm relax} \,=,{{N}\over{{\sqrt{8} \rm ln}\,N}} \,
{r_v^{3/2}\over{G^{1/2}\,N^{1/2}\,\langle m\rangle^{1/2}}} \,\,.
\end{equation}
Dividing equation~(10) by equation~(11) yields
\begin{equation}
{t_b\over{t_{\rm relax}}} \,\sim\, {{10 N\,{\rm ln}\,N}\over f_m^2} 
\left({\langle m\rangle\over m}\right)^{10} \,\,,
\end{equation}
which is a simple modification of the analogous equation~(5).

On the righthand side of equation~(12), the factor $f_m^{-2}$ is necessarily greater than unity. On the other hand, \hbox{$\left(\langle m\rangle/m\right)^{10}$} is, in practice, so small that \hbox{$t_b\,<\,t_b^\ast$}. Consider, for example, the models described in Section~2.2, which had a stellar mass distribution appropriate for the infant Pleiades. Here, \hbox{$\langle m\rangle\,=\,0.36\,\,\Msun$}. In our \hbox{$N\,=\,4096$} cluster, we find empirically that the minimum mass in any newly formed binary is \hbox{$m\,\approx\,4\,\,\Msun$}; the corresponding $f_m$-value is $8\times 10^{-3}$. Equation~(12) then predicts that \hbox{$t_b/t_{\rm relax}\,\sim\,0.2$}, in good agreement with our numerical results. 

This derivation is, of course, highly simplified, and the quantitative result above should not be given too much weight. The relative velocity of an encounter in the core will typically be larger than in the rest of the cluster, and the core density will be higher than the average. A more complete derivation of $t_b$ would also consider the physical basis for the minimum mass $m$.  Presumably, this limit is set by the rate at which dynamical friction allows stars of various mass to drift inward. We will not embellish the argument along these lines, but simply note that equation~(12) adds justification for our main points: (1) The rate of binary formation is very sensitive to the stellar mass distribution, and (2) even in hypothetical clusters composed entirely of single stars, binaries form relatively quickly.  It is only by adopting the extreme assumption that these single stars have identical mass that binary formation can be delayed to the point of core collapse.\footnote{Binary formation is also delayed by stellar mass loss; see Section~4.}

\subsection{Energy Input}

A hard binary that resides within a cluster, no matter how it formed, adds energy to the whole system. The process, like the creation of new pairs, is a three-body interaction. As a result of the encounter, the binary usually tightens and releases energy. This heating accounts for the expansion of both the mass and number shells in Figure~3, for \hbox{$t\,>\,t_b$}. Expansion driven by binaries is global, and differs qualitatively from the dual contraction and expansion seen in the single-mass model (Figure~1).

This difference is also apparent when we view the evolution of the cluster's aggregate energy. First, we need to distinguish {\it internal} and {\it top-level} energies. In the first category is the gravitational binding energy of each binary, and the kinetic energy of both component stars with respect to their center of mass. In the top-level category are the center-of-mass kinetic energies of all bound stellar systems, whether single or multiple, and the gravitational potential energy of this array. Thus, the kinetic energy $K_r$ considered previously was actually a top-level quantity. The cluster's total energy $E_0$ is the sum of the two contributions:
\begin{equation}
E_0 = E_{\rm int} + E_{\rm top} \mbox{.}
\end{equation}
Here, we are ignoring the relatively small amount of energy carried off by escaping stars. In the absence of an external tidal field, $E_0$ remains strictly constant. Binary heating, whether by creation of a new pair or interaction of an existing pair with single stars, lowers $E_{\rm int}$ and transfers the same amount of energy to $E_{\rm top}$.
 
The solid curve in Figure~9 shows the evolution of the top-level energy in the model cluster with a realistic stellar mass distribution (\hbox{$N\,=\,4096$}). Here $E_{\rm top}$ is normalized to $E_i$, its initial value. Since this cluster begins with all single stars, $E_{\rm top}$ and $E_0$ are identical at the start. Upon the formation of the first hard binary at \hbox{$t\,=\,0.37\,\,t_{\rm relax}$}, $E_{\rm top}$ takes a substantial, upward jump. Subsequent jumps occur whenever new hard binaries form, or when existing ones impulsively heat the cluster. As an instructive comparison, the dashed curve in Figure~9 shows $E_{\rm top}$ for the single-mass model described in Section~2.1. The curve is very nearly flat. Despite some weak and transient interactions, no stable, hard binaries form over the span of the simulation. 

One interesting feature of Figure~9 is that the jumps tend to diminish with time. Indeed, $\Delta E_{\rm top}$ scales roughly with $E_0$, where the latter approaches zero as the cluster inflates. To see the origin of this scaling, consider in more detail the energetics of the three-body interaction. The energy released as the binary tightens is shared by that pair and the passing star. Both recoil from the site of the original encounter.\footnote{In some cases, the single star changes places with one of the binary components \citep{he96}. This detail need not concern us.} In our simulations, the binary is lifted to a much higher orbit, but usually does not become unbound. The pair then drifts back down, via dynamical friction, and gives its energy to surrounding stars. If \hbox{$\Delta E_b$} denotes this contribution to the total energy change \hbox{$\Delta E_{\rm top}$}, then \hbox{$\Delta E_b\,\lesssim\,m_b\,\sigma^2$}, where $m_b$ is the binary mass. 
 
The recoiling single star, of mass $\langle m\rangle$, rockets away at high speed, much larger than $\sigma$, and is lost to the cluster. On its way out, the star does work \hbox{$\Delta E_s \,=\,\langle m\rangle\,\Phi_g$} on the system. Here, $\Phi_g$ is the depth of the top-level gravitational potential well at the interaction site, which is close to the cluster center. Now both $\sigma^2$ and $\Phi_g$ are proportional to $E_{\rm top}$ itself. Thus, the total energy change, \hbox{$\Delta E_{\rm top}\,=\,\Delta E_b\,+\,\Delta E_s$}, is also proportional to $E_{\rm top}$.  

The time when the first hard binary appears, \hbox{$t_b\,=\,0.37\,\,t_{\rm relax}$}, is also when the cluster begins to expand (Figure~3). Thus, the binary immediately begins to heat the system through interactions with its neighbors. Eventually, the binary itself is ejected as a result of such an encounter, to be replaced later by another. Over the course of the simulation, a total of 4 hard binaries arise. But a snapshot of the cluster at any time shows it to contain either a single binary or none at all. For example, the flat portion of the energy curve between \hbox{$t\,=\,0.51\,\,t_{\rm relax}$} and \hbox{$0.70\,\,t_{\rm relax}$} represents such a barren period. It is indeed remarkable, as many investigators have noted, how a handful of binaries control the fate of a populous cluster.

In more detail, there is variation of the heating rate with $N$.  Smaller systems experience fewer binary interactions.  On the other hand, each interaction creates a larger $\Delta E_{\rm top}$ relative to $E_0$.  Larger systems have more frequent interactions, with each contributing less relative energy. In the end, the rate of energy input actually varies little, when averaged over a sufficiently long period.

What if the cluster is seeded with many binaries initially?  Figure~10 shows the evolution of the top-level energy for the Pleiades simulation. There are now many binaries even at the start, and thus no initial period of constant $E_{\rm top}$. Remarkably, however, the evolution is quite similar to the case of no primordial binaries. The top-level energy is changed in a few discrete jumps. These few major interactions always involve binary (or triple) systems composed of the few most massive stars (see also \citet{fm96b}).

The important lesson is that only a special subset of binaries strongly influences a cluster's evolution. These are systems which are relatively massive, wide enough to have a significant interaction cross section with other stars, and yet tight enough to be hard. To be sure, the primordial binaries in the Pleiades-like simulation shown in Figure~7 do halt the initial contraction. Relatively little energy input is required to do so. Virtually all primordial binaries are either of too low a mass, or are so tight that they effectively interact as a single system. It is the subsequent coupling of relatively few massive stars that inject much greater energy and principally drive the cluster's expansion.

\subsection{Very Massive Stars}

We have seen how binary heating can dominate a cluster's evolution. For a realistic stellar mass spectrum, the effect begins very quickly, in less than a single relaxation time. Under these circumstances, the cluster is still very far from the point of true core collapse.

Following our Pleiades study (Paper~I), we have set the maximum mass at $m_{\rm max} = 10\,\Msun$.  The reasoning here was that more massive objects would have ionized the parent cloud, allowing the stars to disperse before they could form a bound cluster. In any event, it is instructive when elucidating basic physical principles, to relax this assumption and gauge the effect. We now allow stars in our $N = 4096$ cluster to be drawn from the same mass function as before, but with a nominal upper limit of $m_{\rm max} = 100\,\Msun$. In practice, no star ever realizes this mass; the largest generated is about $60\,\Msun$. Again, there are no primordial binaries.

Based on our earlier arguments, we would expect binary formation to begin even sooner. Indeed this is the case. The first stable, hard binary forms at \hbox{$t_b\,=\,0.18\,\,t_{\rm relax}$}, a factor of 2 earlier in time. Figure~11 shows the evolution of both Lagrangian mass and number shells.  By either measure, the cluster undergoes global expansion at all radii. Not surprisingly, there are detailed differences from the $m_{\rm max} = 10\,\Msun$ case. The apparent early contraction of the innermost mass shells is now {\it entirely} due to mass segregation.  Even the number shell with $N_r / N_0 = 0.03$ expands from the start.

A closer analysis shows that there is again never more than a single binary in the cluster at any instant, although the specific pair changes identity in time. Both components of this pair are always among the top 5 stars by mass. These binaries generate especially strong heating.

Figure~12 shows the evolution of the top-level energy.  For comparison, we also reproduce the analogous plot from Figure~9 for the simulation with $m_{\rm max} = 10\,\Msun$. Binary heating now begins much sooner, and the individual three-body encounters inject larger amounts of energy. This result corroborates our earlier conclusion that $\Delta E_b$ is proportional to the mass of the binary system.

In the presence of very massive stars, the cluster energy, $| E_0 |$, diminishes to only 7~percent of its initial value over the time considered.  In some of our simulations, the heating was so severe as to effectively dissolve the cluster, inflating it to thousands of times its initial size (in the absence of a tidal field). Globular clusters may have been born in parent clouds so massive that even multiple stars producing HII regions do not disrupt them \citep{kb02}. Why, then, are young globular clusters not dispersed by binary heating?  How do they evolve to the point of core collapse? To answer these questions, we now include the last key ingredient - stellar evolution.

\section{The Role of Stellar Evolution}

It has long been appreciated that the mass loss associated with stellar evolution can have a dramatic effect on the early life of a cluster \citep{ag77,a86,t87,b08}.  As its largest stars die out, the cluster's total mass can decrease significantly.  The loss of gravitational binding causes the cluster to expand. This initial phase of expansion, which is ubiquitous in simulations, is quickly stifled because lower mass stars survive much longer.

The loss of the cluster's most massive stars has another effect, more relevant here, that is not as widely appreciated \citep[see however][]{fm96a}. As we have seen, it is these same stars that reverse core contraction and drive global expansion through binary formation and heating. Because of mass loss, however, the objects die out before they can pair with others. Stellar evolution thus tamps down binary heating and postpones the global expansion that this heating drives.

The code Starlab is able to track stellar evolution, including mass loss, by applying analytic fitting formulae. Once we switch on this module, however, we need to give an explicit size scale for our cluster, in order to set the relation between dynamical and stellar evolutionary times.  We select a virial radius of $r_v = 4$~pc as a representative value. In our stellar mass function, we continue to set $m_{\rm max} = 100$~M$_\sun$. As before, we focus on the case $N = 4096$. Our cluster has an initial crossing time of 8~Myr, and an initial relaxation time of $t_{\rm relax} = 570$~Myr. We follow the evolution of the cluster for 8.5~Gyr, which is about 15 relaxation times.

Figure~13 shows the evolution of Lagrangian radii. At first glance, the pattern looks similar to Figure~11, which shows the same cluster, but without stellar evolution. Closer inspection reveals important differences. The cluster now undergoes a rapid expansion. During the first 800~Myr, corresponding to 1.5 initial relaxation times, the virial radius increases by a factor of 2.  Once the maximum mass of the stars falls below about 2~M$_\sun$, the expansion slows.

The next, slower phase of expansion lasts until about 3~Gyr, or 5.5~$t_{\rm relax}$. Here, heating is provided by an 18~M$_\sun$ black hole left behind by a formerly 67~M$_\sun$ star. Due to its mass, it readily forms a binary system with another star, and this system is the source of the heating. The quantitative details of this phase are as uncertain as our knowledge of the late stages of massive stellar evolution. For example, \citet{h00} find that the same 67~M$_\sun$ star leaves behind a 4~M$_\sun$ black hole, which would create much less heating and expansion.

Neither of these phases are in the previous Figure~11, which omitted stellar evolution. Instead there is an initial brief contraction of the innermost mass shells. As noted earlier, the innermost number shells do not contract, so we are actually witnessing the effects of mass segregation. In the present case, significant contraction of both the mass and number shells occurs. The 18~M$_\sun$ black hole and its companions have been ejected, and no new binaries form.  Hence the system is undergoing true dynamical relaxation. Compared to the system with no stellar evolution, this phase is quite protracted, lasting 6~$t_{\rm relax}$.\footnote{Due to the earlier expansion, the relaxation time at the start of core contraction is 10 times longer than its initial value. The contraction phase lasts for only 0.6 times this readjusted and more appropriate relaxation time.} Again, the most massive binaries, that would have halted contraction earlier have died off.

Eventually, however, new binaries do form.  As before, it is the highest mass stars present that interact and cause heating. The cluster thereafter enters a prolonged phase of global expansion.  This lasts through the end of the simulation.  In summary, stellar mass loss has delayed binary formation, and therefore cluster expansion, but not prevented their occurrence.

\section{Discussion}

\subsection{Open vs. Globular Clusters}

The simulations we performed without stellar evolution all found that $t_b$, the epoch marking the onset of binary formation, was a fixed fraction of $t_{\rm relax}$.  We have just seen, in the specific case of $N = 4096$ that stellar mass loss modifies this result, increasing $t_b / t_{\rm relax}$. Another path to the same conclusion comes from equation~12.  Stellar evolution lowers the minimum mass $m$ of stars that are around to form a hard binary that can heat the cluster.  The ratio $\langle m \rangle / m$ thus increases, and $t_b / t_{\rm relax}$ rises accordingly.

As we consider clusters of higher $N$, a basic point to note is that $t_{\rm relax}$ itself increases.  If the average mass $\langle m \rangle$ is unchanged, then equation~11 shows that $t_{\rm relax}$ scales as $N^{1/2} r_v^{3/2}$, ignoring the logarithmic factor. Thus, for similar virial radii $r_v$, clusters of higher population take longer to relax. The binary formation time $t_b$ in these systems is longer, and, because of stellar evolution, is a higher fraction of $t_{\rm relax}$ itself.

By a given age, therefore, a cluster of higher population has experienced more dynamical relaxation.  That is, its core has contracted further. In our view, this trend represents the critical difference between open and globular clusters.  The former undergo, at best, a brief, tepid period of core contraction. In our Pleiades simulation, even this mild contraction is stifled by heating from primordial binaries. Globular clusters, on the other hand, undergo much longer periods of dynamical relaxation. In some cases, this prolonged epoch results in true core collapse.

\subsection{Cluster Death}

Throughout this study, we have carried out our simulations to arbitrary times, just long enough to illustrate the main evolutionary phases.  The smaller-$N$ groups on which we focus, are eventually destroyed tidally, either by the general Galactic field or by the close passage of giant molecular clouds.  \citet[][equation~(8.57)]{bt08} give the cloud disruption time scale as
\begin{equation}
t_{\rm dis} = 250 \mbox{ Myr} \left( \frac{M}{300 \mbox{ M}_\sun} \right)^{1/2} \left( \frac{r_h}{2 \mbox{ pc}} \right)^{-3/2} \mbox{,}
\end{equation}
where $r_h$ is the half-mass radius. Consider again our simulation of an $N = 4096$ cluster, with stellar evolution included. Here, $M = 1700$~M$_\sun$ and $r_h = 3.3$~pc. According to equation~(14) $t_{\rm dis} = 280$~Myr, or $0.5 t_{\rm relax}$. Figure~13 shows that the cluster is torn apart very early, during the initial phase of rapid expansion accompanying the death of its most massive stars.

Tidal disruption by passing clouds has long been considered the dominant cluster disruption mechanism \citep{s58}. As noted, however, even the Galactic tidal field will eventually do the job.  In our Pleiades simulation of Paper~I the cluster was largely destroyed in this way by 700~Myr. An isolated cluster of this size would be just entering its phase of weak contraction. Because of this external tidal field, however, the simulated Pleiades never even began to contract, but globally expanded until it totally dissolved. Such tidal disruption may account for the mass-independent pattern of cluster death observed in the Antennae Galaxies \citep{f09} as well as the the Magellanic Clouds \citep{c10}.

A few open clusters do survive for ages much longer than the ones just mentioned \citep{f95}. These lie in the outer reaches of the Galaxy, where the encounter rate with giant molecular clouds is relatively low, and the general tidal field is also weaker. Our isolated $N = 4096$ cluster eventually begins core contraction at 3~Gyr, corresponding to $6\,t_{\rm relax}$. Even the weakened Galactic tidal field will begin to disrupt the system by this age, as seen in the simulation of the even richer system M67 \citep{h05}. The lesson here is that, while open clusters can in principle enter a phase of tepid core contraction, none reach this point in reality.

\subsection{Summary}

The classical theory of dynamical relaxation is relatively simple, an elegant illustration of how systems with negative heat capacity evolve \citep{l99}. However the theory does not describe accurately real clusters, at least those of modest population on which we have focused.  The two main factors changing the picture are binary heating and stellar evolution. Both processes are, of course, well understood, but their combined effect has not been appreciated.

All clusters are born with a large fraction of binaries, but these do not provide the largest effect. It is the system's most massive stars coupling together that generate most of the heating through three-body interactions. This heating easily reverses incipient core contraction, so that the central density climbs only slightly before the new phase of global expansion begins. This phase resembles, at least qualitatively, the post-collapse evolution described by \citet{h72}. However, the reversal from contraction occurs at much lower density than in earlier accounts.

Mass loss accompanying stellar evolution modifies the picture, but does not change it qualitatively. Since the most massive stars die out before they can couple with others, the degree of binary heating, and therefore the vigor of global expansion, is less. In addition, the earlier phase of core contraction lasts longer and leads to a higher central density before reversal. Both modifications increase with the cluster population $N$. We thus see why some globular clusters indeed reach the point of true core collapse, which can be reversed only by the tightest of binaries.

The new picture of cluster evolution presented here is more complex than the classical one, but it is motivated by the basic physical effects that are incorporated in modern numerical simulations. With the benefit of hindsight, it is easy to see why earlier, simplified methods reinforced the impression that dynamical relaxation is ubiquitous. In single mass models, binary formation is so delayed that it becomes irrelevant. Statistical models, based on solving the Fokker-Planck equation, neglect three-body effects entirely. Finally, the contraction of Lagrangian mass shells is not a reliable sign of core contraction, but may reflect a different phenomenon, mass segregation. Our new picture is itself far from complete. Future simulations carried out at higher $N$ will reveal in detail how the transition is made to a more vigorously contracting central core.

\acknowledgments
We are grateful to Douglas Heggie and Simon Portegies-Zwart for helping us navigate the literature of dynamical relaxation. Onsi Fakhouri also provided useful suggestions on the visualization of energy transfer. This research was supported by NSF grant~AST~0908573.

\clearpage




\clearpage

\begin{figure}
\plotone{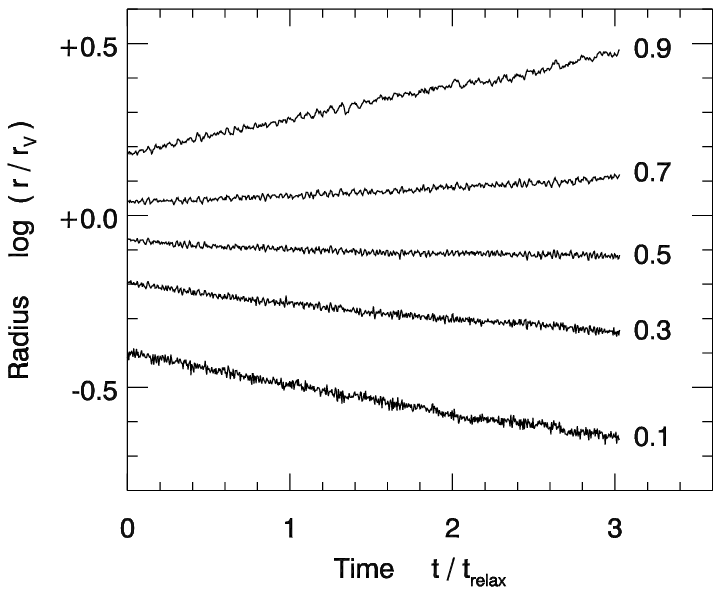}
\caption{Temporal evolution of the radii of Lagrangian mass shells, for a single-mass cluster model (\hbox{$N\,=\,4096$}). Each curve is labeled by the corresponding mass fraction of the cluster. The radii are normalized to the initial virial value, $r_v$, and the time to the initial relaxation time, $t_{\rm relax}$.}
\end{figure}

\clearpage

\begin{figure}
\plotone{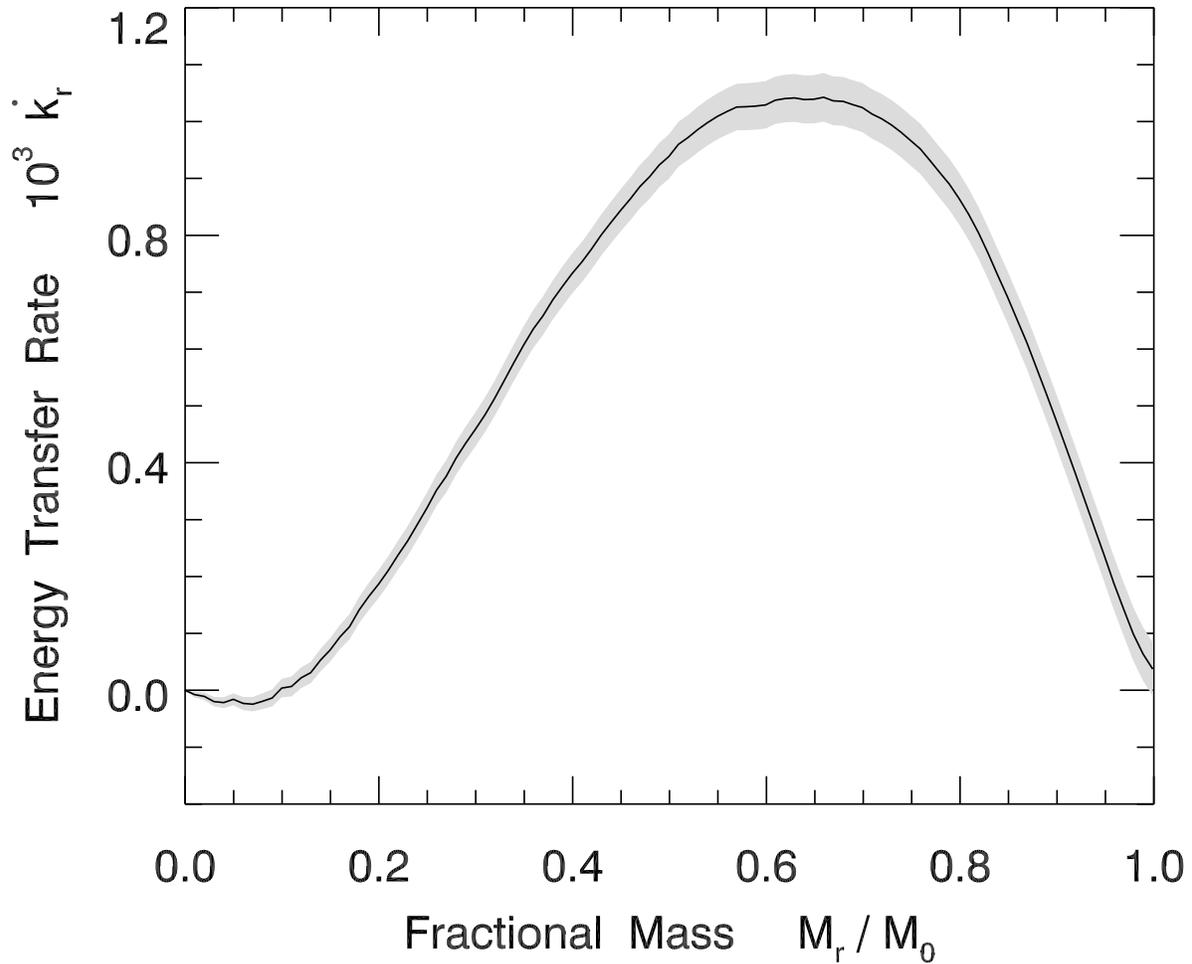}
\caption{Profile of the mean energy transfer rate for a single-mass cluster model ($N = 4096$). The rate, displayed as the nondimensional quantity $\dot{k}_r$ given in the text, is plotted against the mass fraction $M_r/M_0$. The shading indicates the 1-$\sigma$ error in the value of $\dot{k}_r$ at each point.}
\end{figure}

\clearpage

\begin{figure}
\plotone{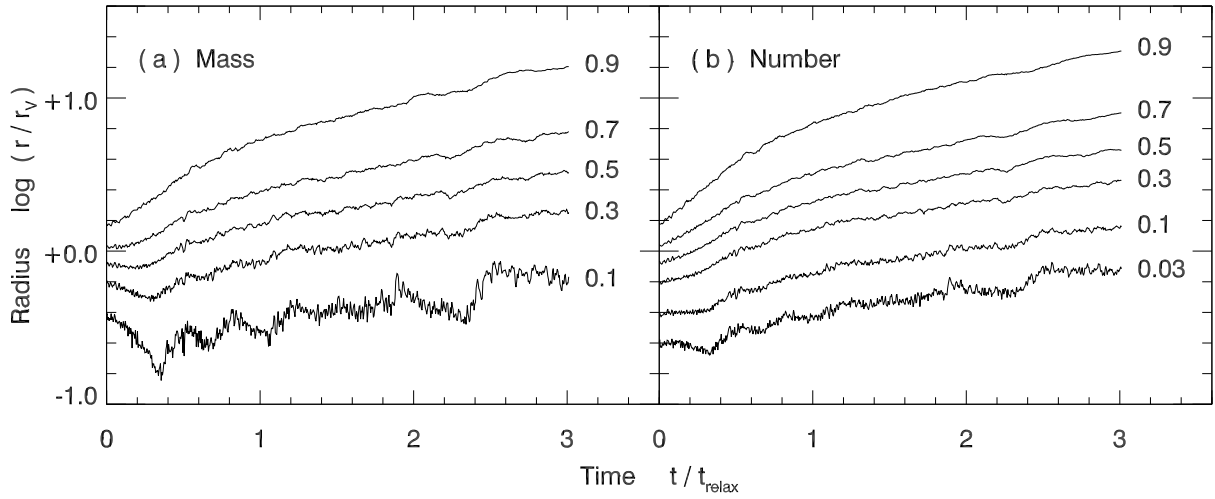}
\caption{Evolution of the radii of (a) mass shells, and (b) number shells, for a cluster with a realistic stellar mass distribution (\hbox{$N\,=\,4096$}). Each curve is labeled by the appropriate mass or number fraction of the entire cluster.}  
\end{figure}

\clearpage

\begin{figure}
\plotone{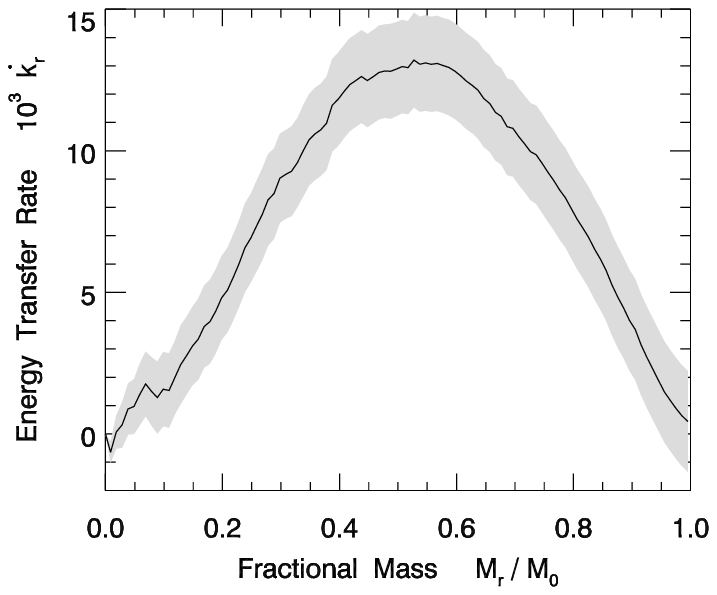}
\caption{Early-time energy transfer profile for an $N = 4096$ cluster with a realistic stellar mass distribution. As in Figure~2, the shading indicates the estimated uncertainty at each point.}
\end{figure}

\clearpage

\begin{figure}
\plotone{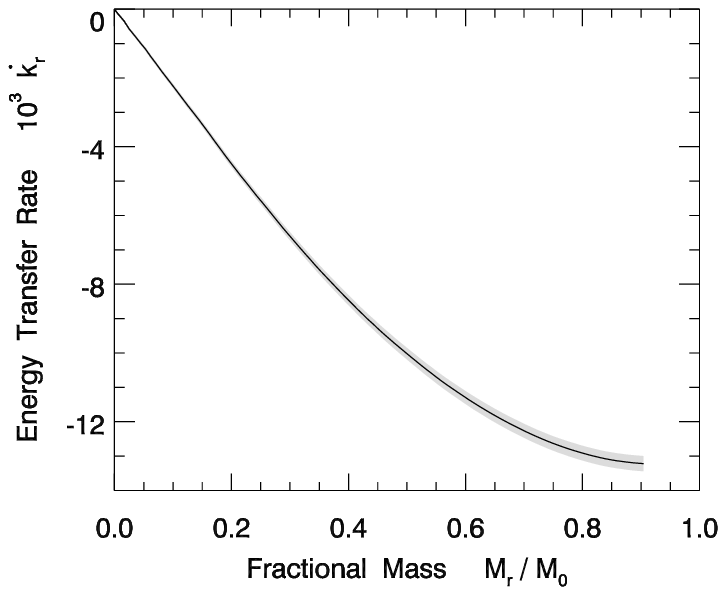}
\caption{Late-time energy transfer profile for an $N = 4096$ cluster with a realistic stellar mass distribution. As in Figure~2, the shading indicates the estimated uncertainty at each point.}
\end{figure}

\clearpage

\begin{figure}
\plotone{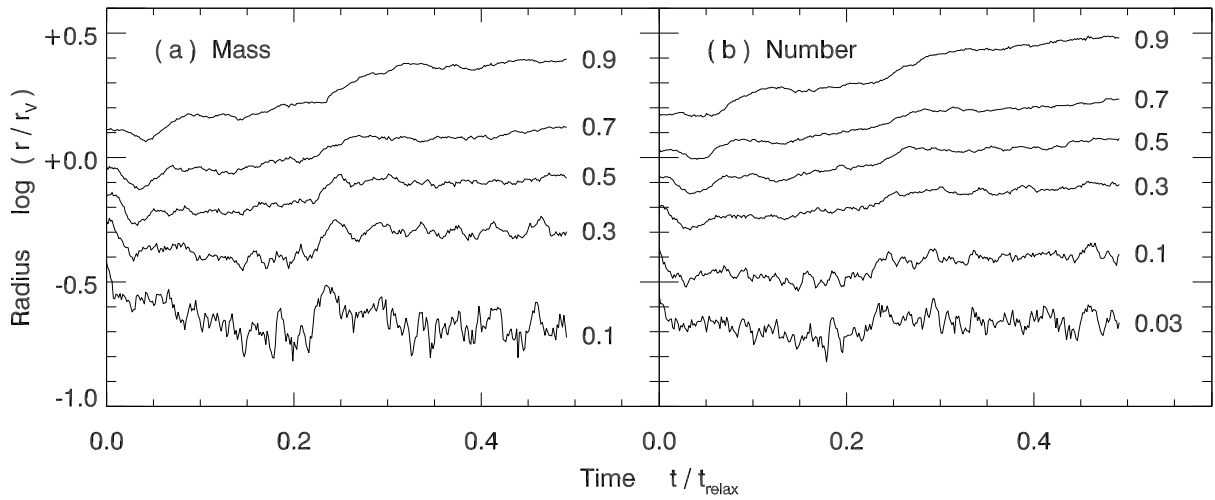}
\caption{Evolution of the radii of (a) mass shells, and (b) number shells, for the Pleiades (\hbox{$N\,=\,1215$}). As in Figure~3, which covers a much longer time, each curve is labeled by the appropriate mass or number fraction.}
\end{figure}

\clearpage

\begin{figure}
\plotone{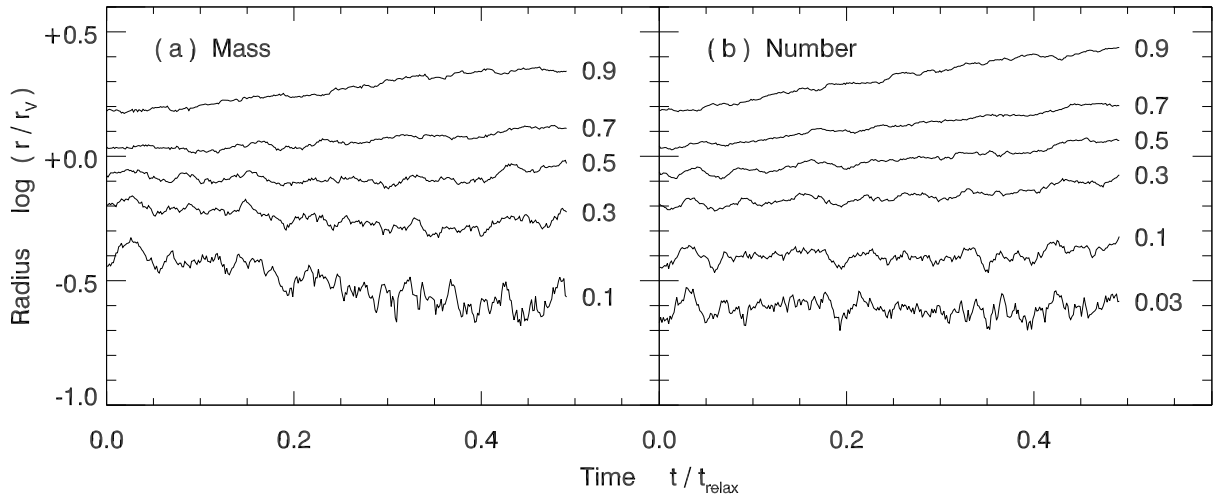}
\caption{Evolution of the radii of (a) mass shells, and (b) number shells, for a Pleiades-like cluster with no initial mass segregation.}
\end{figure}

\clearpage

\begin{figure}
\plotone{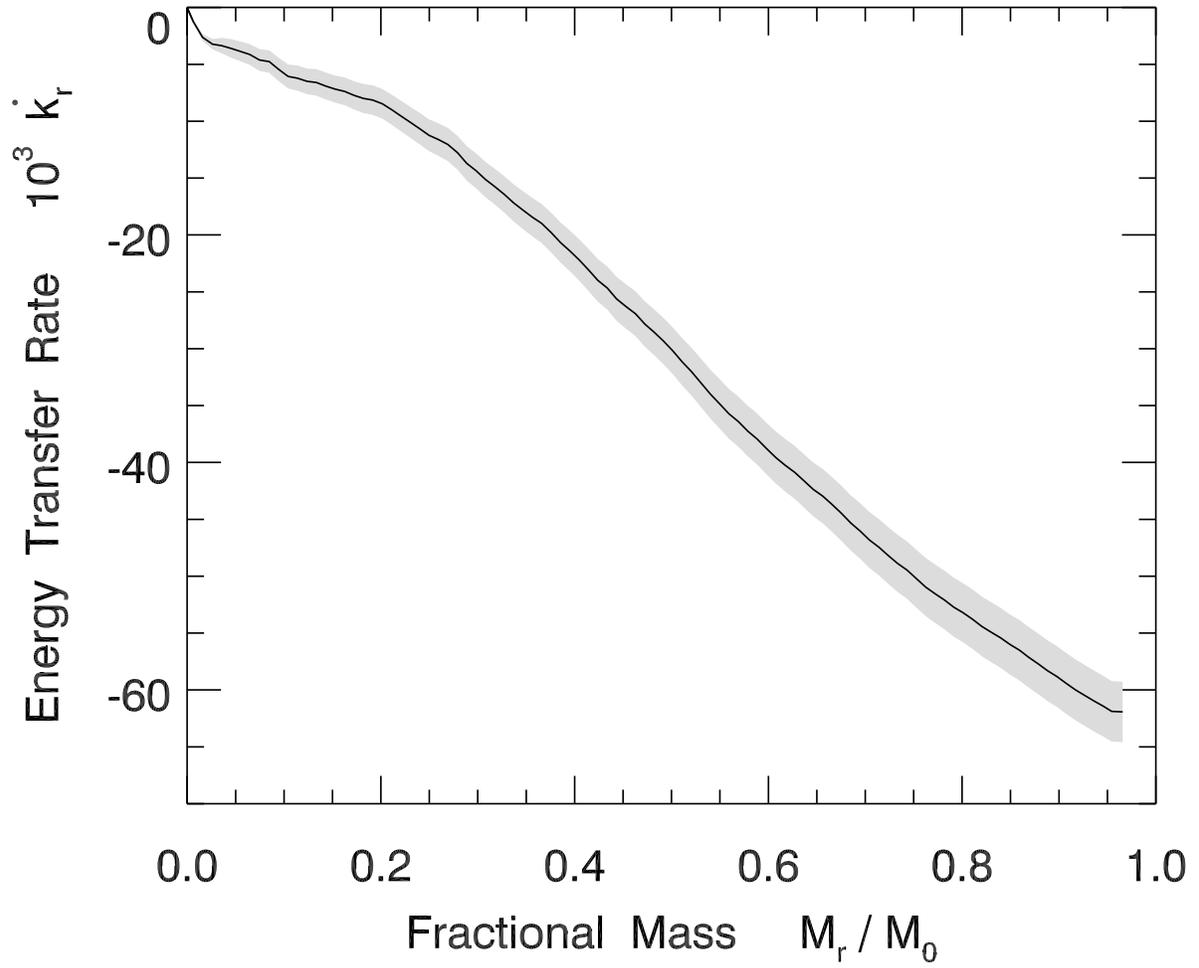}
\caption{Energy transfer profile for the Pleiades. As in Figure~2, the shading indicates the estimated uncertainty at each point.}
\end{figure}

\clearpage

\begin{figure}
\plotone{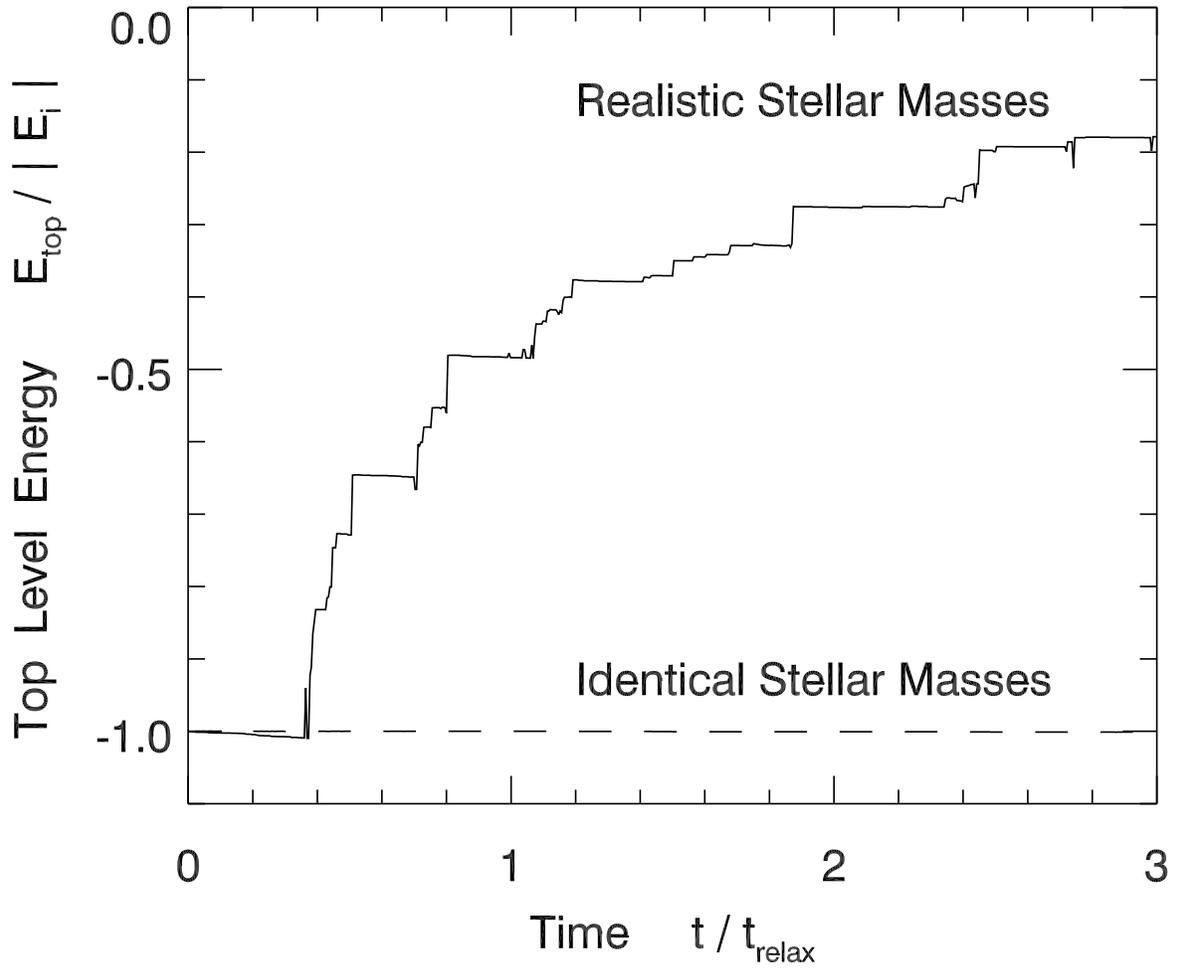}
\caption{Energy evolution for an $N = 4096$ cluster with a realistic mass distribution. The dashed curve shows the evolution in the analogous, single-mass model.}
\end{figure}

\clearpage

\begin{figure}
\plotone{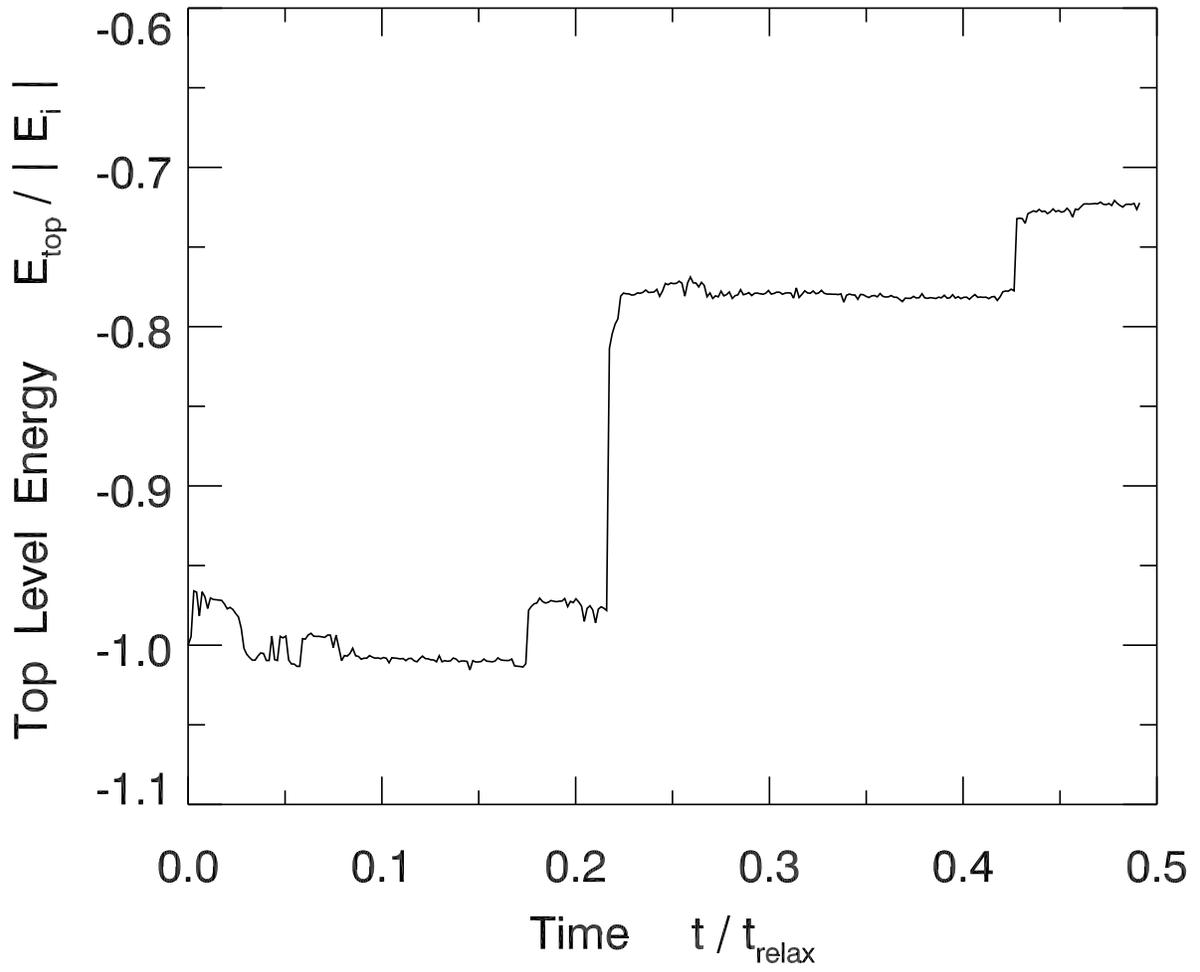}
\caption{Energy evolution for the Pleiades.}
\end{figure}

\clearpage

\begin{figure}
\plotone{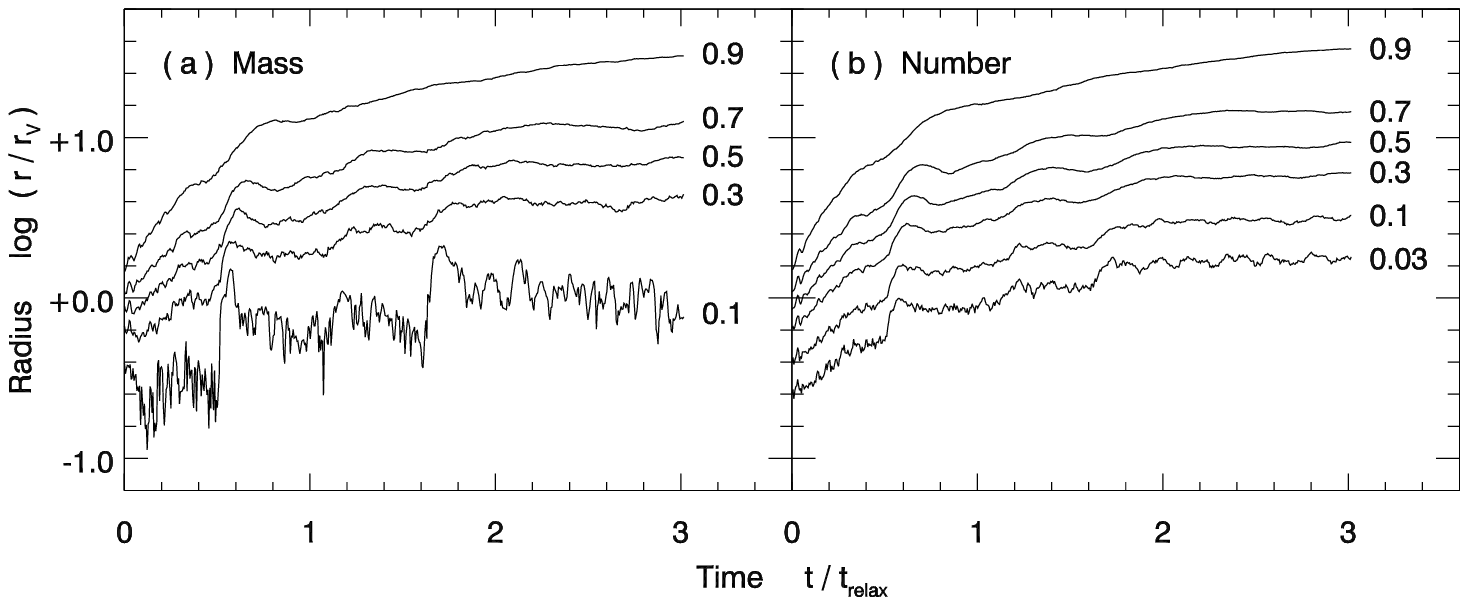}
\caption{Evolution of the radii of (a) mass shells, and (b) number shells, for an $N = 4096$ cluster that includes very massive stars. As in Figure~3, each curve is labeled by the appropriate mass or number fraction.}
\end{figure}

\clearpage

\begin{figure}
\plotone{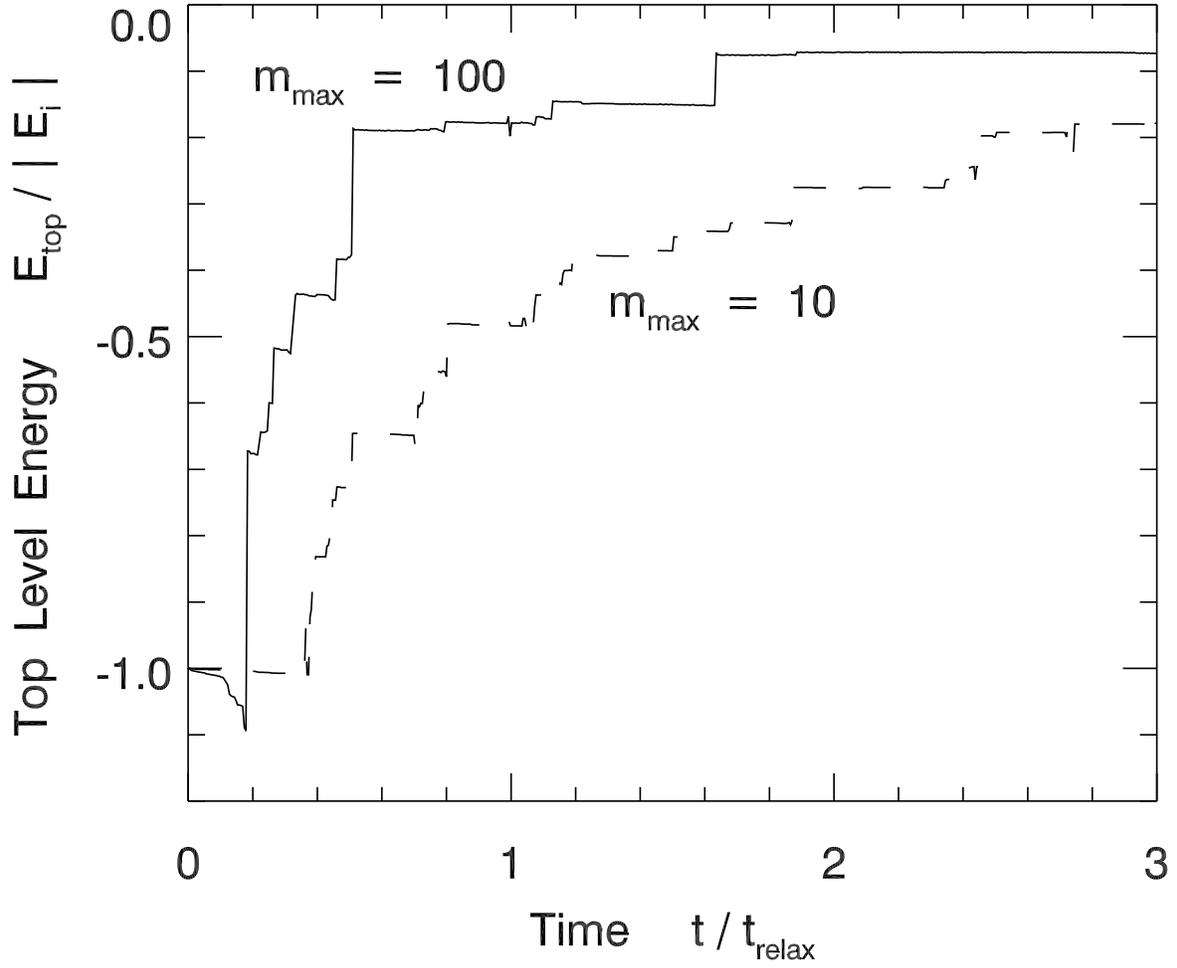}
\caption{Energy evolution for an $N = 4096$ cluster that includes very massive stars.  For comparison, the dashed curve reprodcues that from Figure~9, where the maximum mass is 10 M$_\sun$.}
\end{figure}

\clearpage

\begin{figure}
\plotone{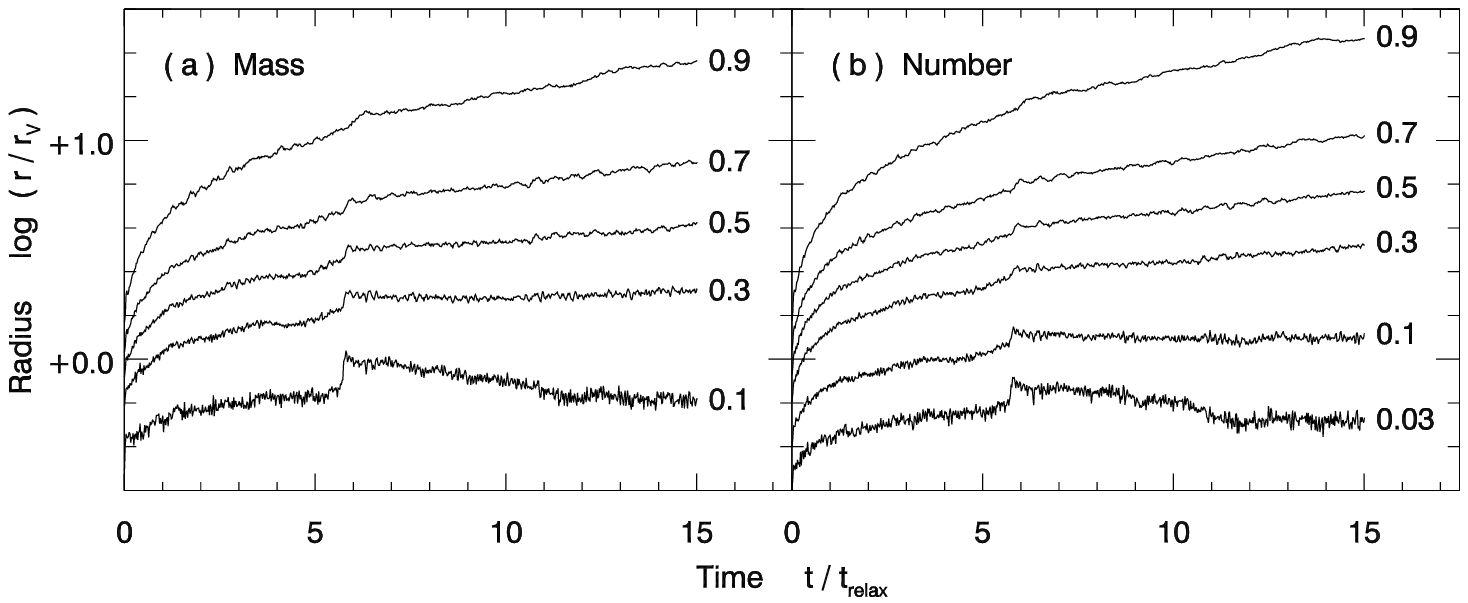}
\caption{Evolution of the radii of (a) mass shells, and (b) number shells, for an $N = 4096$ cluster that includes stellar mass loss. As in Figure~3, each curve is labeled by the appropriate mass or number fraction.}
\end{figure}


\end{document}